\documentclass[11pt]{article}
\usepackage{graphicx}
\usepackage{amsmath}
\usepackage{amssymb}
\usepackage{amsxtra}

\title{Dusty Dark Matter Pearls Developed. }
\author{H.B. Nielsen\footnote{Speaker at the  Work Shop
    ``What comes beyond the Standard Models'' in Bled.}\\ Niels Bohr Institut\\
  hbech@nbi.dk\\
  C.D.Froggatt\\ Glasgow University\\Colin.Froggatt@glasgow.ac.uk }
\date{``Bled''   , July  2022}

\begin{document}
\maketitle

\begin{abstract}
  We briefly review and update our earlier published model for dark matter consisting of nanometer
  size bubbles of a new speculated vacuum phase in which some ordinary material,
  e.g. carbon, is present under high pressure caused by the surface tension of
  the domain wall surrounding the bubble. These bubbles or pearls are surrounded by dust grains, and it is one of the new points of the present article that
  this dust grain rather than being a three-dimensional blob of ordinary matter
  is a lower dimensional object of some lower Hausdorff dimension. We make
  several order of magnitude fits and find rather good agreement for our model.
  However we must imagine that the very high  - 3.5 kev - homolumo gap assumed present in the highly compressed medium inside the bubble has influenced the neighbouring dust, so as to make it significantly harder than usual dust. This is to ensure that we obtain the correct order of magnitude for the velocity in the collisions between dark matter particles at which the cross section falls strongly with increasing velocity.  
 
  The dark matter pearls lose their surrounding dust in passing through the earth's atmosphere and impacting the earth. They must reach down with their terminal velocity through the shielding to the DAMA observatory in less than a year, so that there is no problem in obtaining the seasonal variation effect the DAMA experiment has observed. This is easier to achieve with a sufficiently high pearl mass and this mass is not so strongly restricted once the dust grain is lower dimensional.   
  
\end{abstract}



\section{Introduction
}\label{s:intro}

What dark matter really consists of is one of greatest mysteries in physics
today, and we have long worked on the proposal that it consists of bubbles
of a new phase of the vacuum into which is filled some ordinary material, such
as probably carbon in the form of highly compressed diamond
\cite{Dark1, Dark2, Tunguska, supernova, Corfu2017, Corfu2019, theline,
  Bled20, Bled21, extension, Corfu21}. Our main assumption not based simply
on the Standard Model is that there are several possible phases of the vacuum
with {\em the same energy density} \cite{MPP1, MPP2, MPP3, MPP4, tophiggs,
  Corfu1995}. So it is only the surface tension $S$ of the surface
between the ``new'' vacuum inside the bubble and the usual vacuum outside the
bubbles which keeps the diamond under high pressure. Apart from our new
speculation of there existing several phases of the vacuum, a speculation
with the help of which we {\em pre}dicted the mass of the Higgs boson
\cite{tophiggs} before it was found, our model is based only on the Standard
Model, an achievement not usually managed by models for the dark matter.

At the present, in spite of the gravitational force from the dark matter
fitting the motions of the stars and galaxies and
cosmology well, the remarkable facts are that

\begin{itemize}
\item The majority of the underground experiments - in particular the Xenon
  ones \cite{LUX, Panda, XENON} - do
  {\em not} see any dark matter hitting the Earth, and
\item Accelerators - LHC is the most hopeful - have {\em not} been able to
  see any dark matter either.
  \end{itemize}

   %


\section{Pearl}
   The dark matter particles or pearls are composed of:    
   \begin{itemize}
    \item A nm-size bubble of a new speculated vacuum filled with
       highly compressed atomic stuff, say carbon.
     \item A surrounding dust particle of ``metallicity'' material
       \cite{density, met}
      such as C, O, Si, Fe, ..., presumably of some non-integer
      Hausdorff dimension about 2 or 1. This atomic matter is influenced
      by the electrons being partly in a superposition of being inside
      the bubble of the new vacuum, where there is very high homolumo gap
      between filled and unfilled electron states \cite{Corfu21}. This
      influences the dust grain material so as to make it denser and harder.
    \end{itemize}
  Pearls interact with:
    \begin{itemize}
    \item other dark matter particles and thereby provide an example of
      self-interacting dark matter (SIDM) \cite{firstSIDM}.
    \item atomic matter.
      \end{itemize}

\begin{figure}
\includegraphics[scale=3]{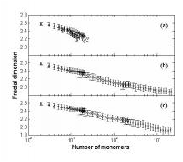}
      
      \includegraphics[scale=3]{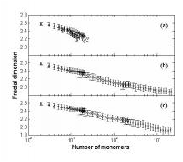}
      \caption{\label{low-dimension} Fractal dimensions of dust grain given
        by the Hausdorff dimension from \cite{Hd}. }
    \end{figure}

  
    The most important evidence for our model may be that we find the
    energy value of 3.5 keV in {\em three} different places as a possible
    favourite energy level difference for dark matter:
    \begin{itemize}
    \item From places in outer space with a lot of dark matter, galaxy
      clusters, Andromeda and the Milky Way Center, as the energy of an
      unexpected X-ray line \cite{Bulbul, Boyarsky, Boyarsky2, Bhargava,
        Sicilian, Foster}, and strangely also from Tycho supernova remnant
      \cite{Jeltema}.
    \item As the average energy of the DAMA dark matter events \cite{DAMA1,
      DAMA2}.
    \item As an average energy for the electron recoil excess\footnote{The
      results from the more sensitive XENONnT detector were
      published \cite{XENONnT} shortly after this School. XENONnT reduces
      the low energy electron recoil background to a factor of ~ 5 lower than
      in XENON1T and  observe no electron excess above background.} in the
      XENON1T experiment \cite{Xenon1Texcess}.
    \end{itemize}
%
    Dust easily gets of lower than 3 dimensions because the growing of a
    dust grain takes place by molecules (monomers) almost one by one being
    attached to the grain as it is at the time and by grains colliding and
    sticking together. Such growing could easily make the dimension
    non-integer. This idea of a dust grain with a fractal dimension has been
    studied in \cite{Hd} and results from this paper are given in
    Figure \ref{low-dimension}. 

\begin{figure}
      \includegraphics[scale=0.5]{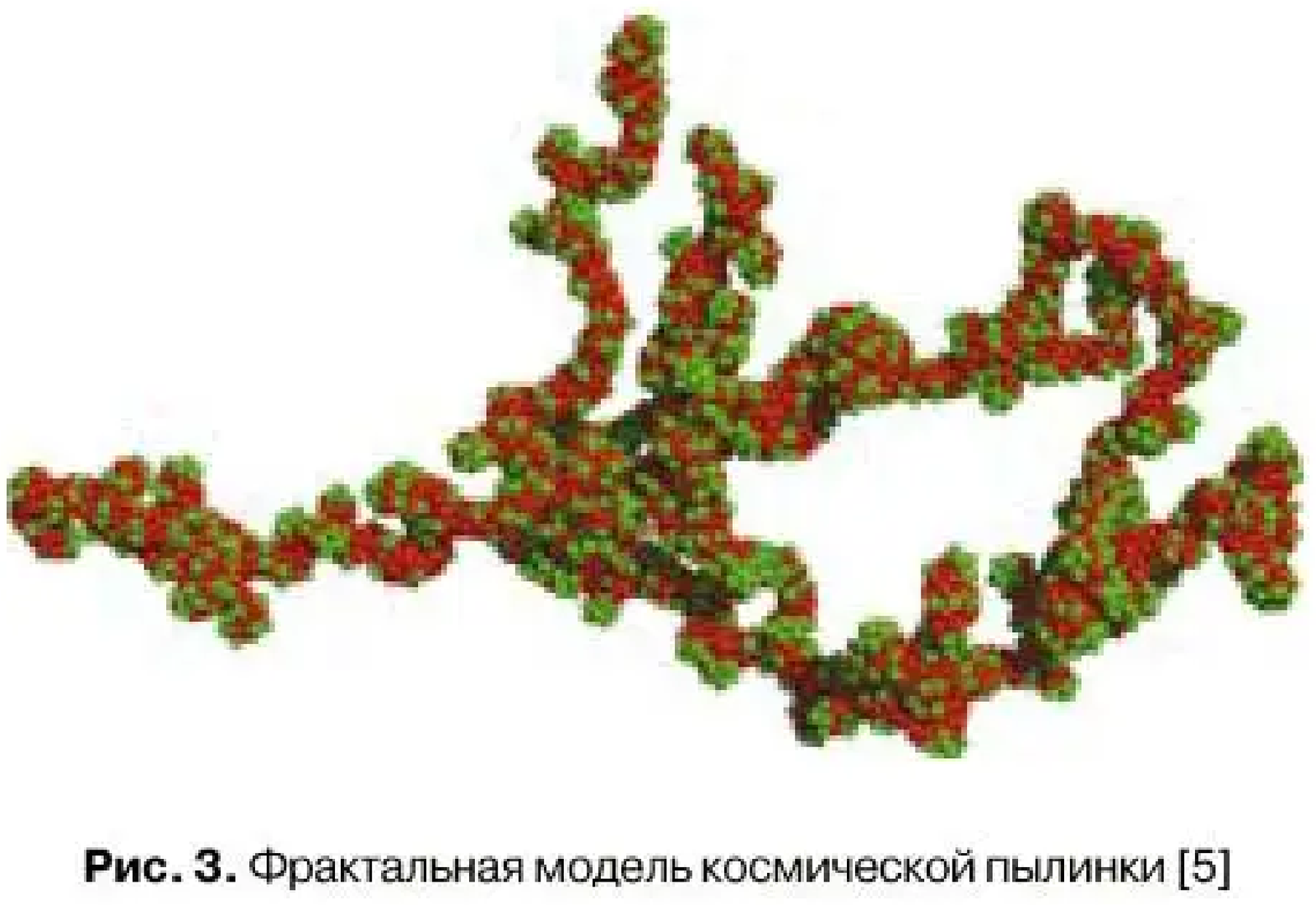}
      \caption{\label{fractal-grain} Picture of Fractal Cosmic Dust Grain
        constructed from 1024 monomers \cite{Wright}.}
      \end{figure}

      An example of such a fractal cosmic dust grain built up from 1024
      monomers \cite{Wright} is given in Figure \ref{fractal-grain}.
      
\begin{figure}
      \includegraphics[scale=0.3]{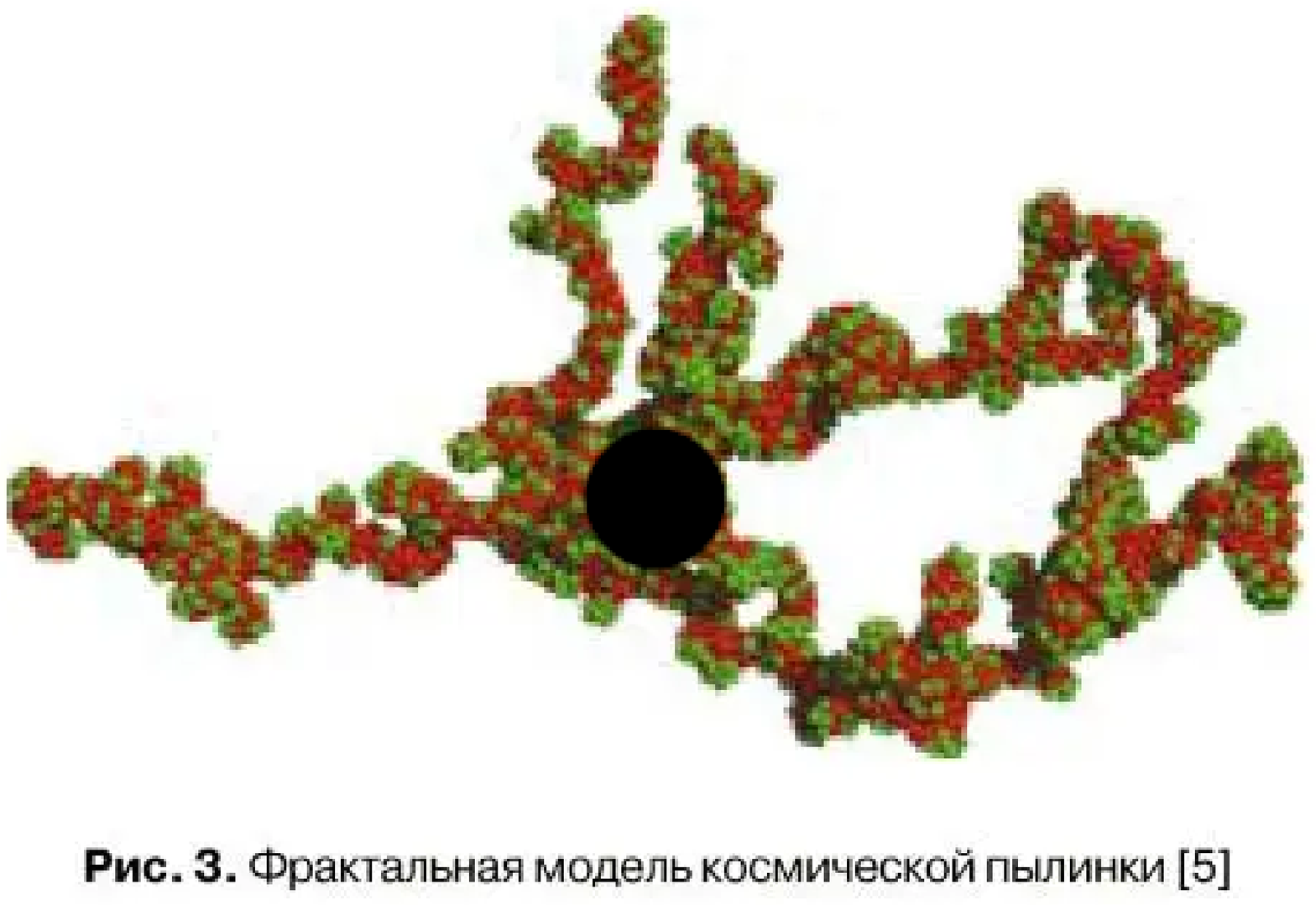}
      \caption{\label{dusty-bubble} The little dot inserted in the foregoing
        figure \ref{fractal-grain} here symbolizes the bubble of new vacuum.
        It is supposedly much heavier than the rest of the dust grain.}
      \end{figure}

      It is indeed very likely that such a dust grain would collect on top
      of one of our pearls, which in itself is very much like a seed atom. We
      may illustrate that in Figure \ref{dusty-bubble} by drawing our little
      pearl as a bubble of new vacuum inside the dust grain.

      If typically the grain size is  $0.1 \mu m$ and the bubble size is
      $1 nm = 0.001 \mu m$, then the bubble is about 100 times smaller than
      the dust grain.
    
\section{Achievements}

 Important achievements of our model:

      \begin{itemize}
      \item Explain that only DAMA ``sees'' the dark matter by the
        particles interacting so strongly as to be quite slow and unable to
        knock nuclei so as to make observable signals. Instead
        the DAMA signal is explained as due to  emission of electrons from
        pearls in an excited state
        with the ``remarkable 3.5 keV energy''.

        Actually Xenon1T may have seen these electrons from the dark matter
        particle {\em decays} as the mysterious electron recoil excess.
      \item The favourite frequency of electron or photon emission
        of the dark matter particles is due to a homolumo gap in the
        material inside the bubble of the new vacuum. This gap should be
        equal to the 3.5 keV. 
        \end{itemize}

      
      \begin{itemize}
      \item We have made a rather complicated calculation of what happens
        when the
        bubbles - making up the main part of the dark matter particle - hit
        each other and the surface/skin/domain wall contract and how one gets
        out a part of the energy as 3.5 keV X-rays \cite{theline}. We fit with
        one
        parameter both the very frequency 3.5 keV, and the
        over all intensity of the corresponding X-ray line observed from
        galaxy clusters etc. This production mechanism gives an intensity
        proportional to the dark matter density squared and we use the results
        of the analysis of Cline and Frey \cite{FreyCline} whose model shares
        this property.
      \item We explain why - otherwise mysteriously - the 3.5 keV line
        was seen by Jeltema and Profumo \cite{Jeltema} from the Tycho
        supernova remnant and probably also problems with the Perseus galaxy
        cluster 3.5 keV
        X-ray observations. This is by claiming the excitation of the bubbles
        come from cosmic radiation in the supernova remnant.
        \end{itemize}

\begin{figure}
  \includegraphics[scale=0.65]{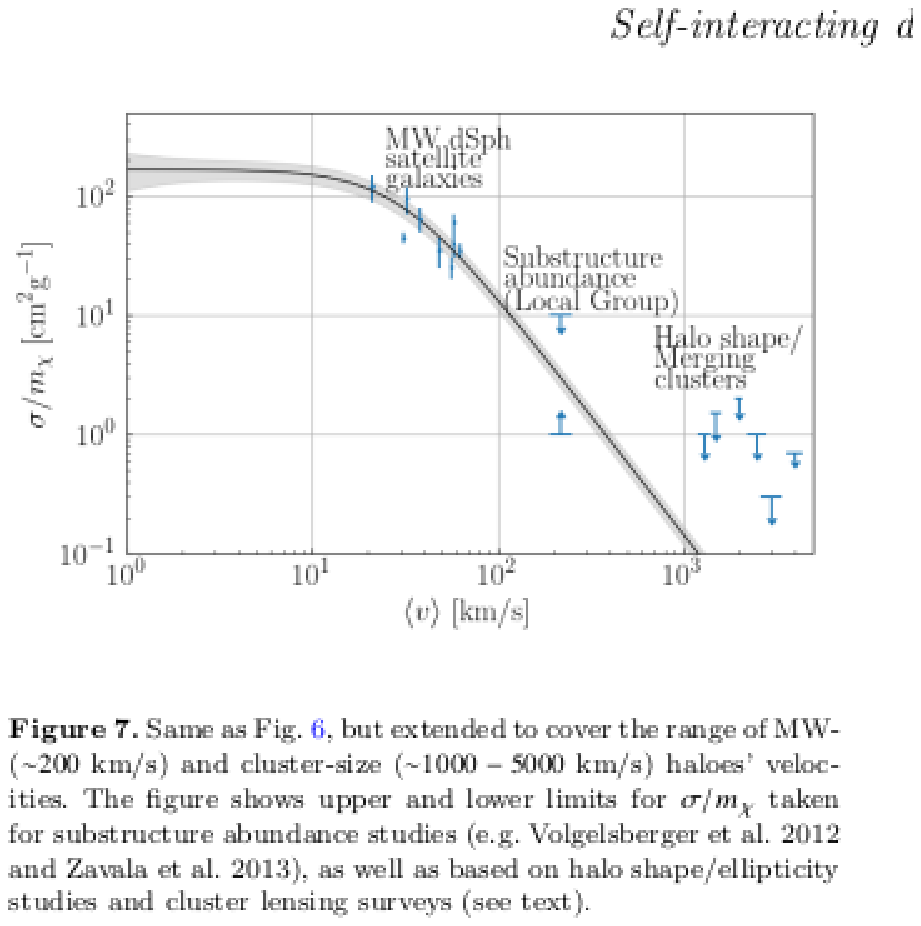}
  \caption{\label{Correa} Extract from Correa’s paper illustrating the fits
    of the cross section to mass ratio obtained for
    different dwarf galaxies, as a function of the estimated velocity
    for the relevant galaxy.} 
  \end{figure}

      
      \begin{itemize}
      \item According to expectations from ideal dark matter that only
        interacts essentially by gravity there should be e.g in a dwarf galaxy
        a concentrated peak or cusp of dark matter, but that seems not to be
        true. The inner density profile rather seems to be flat as expected for 
        self-interacting dark matter \cite{firstSIDM}.
        Correa \cite{CAC} can fit the dwarf galaxy star velocities by the
        hypothesis that
        dark matter particles interact with each other with a cross section
        over mass ratio increasing for lower velocity, as shown in
        Figure \ref{Correa}. We fit the cross section
        over mass velocity dependence of hers. But we need a ``hardening ''
        of the dust around the bubbles.
      \end{itemize}
    
    \section{Impact}
    

  \begin{figure}
    \includegraphics[scale=0.6]{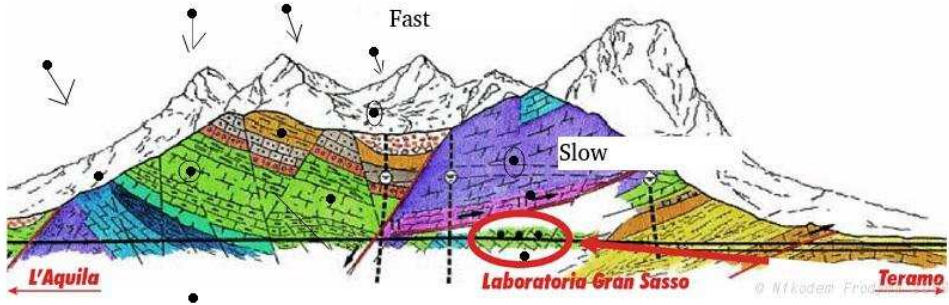}
    \caption{\label{ip} The mountains above the Gran Sasso laboratories.}
  \end{figure}

  
      We imagine that the dark matter pearls lose their dust grain in the
      atmosphere or at least, if not before, by the penetration into the earth
      shielding, and that they at the same time get excited by means of the
      energy from the braking of the pearls. For a very small number of the
      pearls this excitation energy gets radiated first much
      later when the pearl has passed through the earth shielding to the
      underground detectors, so as to
      deliver X-ray radiation with just the characteristic 3.5 keV energy
      per photon. The energy is delivered we guess by electrons or photons.
      Thus experiments like the xenon experiments do not ``see'' it
      when looking for nucleus-caused events. Only DAMA, which does
      not notice if it is from nuclei or from electrons, does not throw
      electron-caused events away as something else.
  
  The dark matter pearls come in with high speed (galactic velocity), but
  get stopped down to much lower speed by interaction with the shielding
  mountains, whereby they also get excited to emit 3.5 keV X-rays or
  {\em electrons}.

\section{Calculations}
  
  The percentages of the matter in the Universe are as follows:

  \begin{itemize}
  \item  27\% dark matter (while  68\% of a form of energy known as dark
    energy, and 5 \% ordinary matter).
  
     \end{itemize}
  The elements heavier than hydrogen and helium make up of order 1\% of
  ordinary matter and are known as ``metals". The comoving density of
  these ``metals" together is \cite{density}
   \begin{eqnarray}
   	\hbox{``metal density"} &=& 5.48*10^6 M_{\odot}Mpc^{-3} \\
   	 &=& 3.71*10^{-31} kg/m^3.
   \end{eqnarray}
\subsection{Inverse Darkness $\frac{\sigma}{M}$}
We think that the less the cross section is compared to the mass the less
is the interaction - with whatever we may consider - and thus the less the
``visibility". It is the lack of visibility which we call darkness just as
we call dark matter dark matter because we do not ``see" it. So a smaller
cross section means darker and thus the ratio with the cross section in the
numerator could be called the inverse darkness.

Using the atomic radii we can calculate the cross sections for the
following atoms:

  \begin{eqnarray}
  \hbox{Hydrogen H: } r_H &=& 25 pm\Rightarrow \sigma_{H}=\pi r_H^2 =1963 pm^2 \\
  \hbox{Helium He: } r_{He} &=& 30 pm\Rightarrow \sigma_{He}=\pi*r_{He}^2=
  2827pm^2\\
  \hbox{Carbon C: } r_C &=& 70 pm \Rightarrow \sigma_C=\pi*r_C^2=15394pm^2 \\
  \hbox{ Silicium Si: } r_{Si} &=& 110 pm \Rightarrow \sigma_{Si}=\pi*r_{Si}^2=
  38013 pm^2
  \end{eqnarray}

  
Using that one atomic unit $1 u$ = $1.66*10^{-27}kg$ we get for the inverse
darkness ratios for the atoms mentioned:

\begin{eqnarray}
  \hbox{Hydrogen H: } \frac{\sigma_H}{1u*1.66*10^{-27}kg/u}&=& 1.183*10^6m^2/kg\\
  \hbox{Helium He: }\frac{\sigma_{He}}{4u*1.66*10^{-27}kg/u}&=& 4.26*10^5m^2/kg\\
  \hbox{Carbon C: }\frac{\sigma_C}{12u*1.66*10^{-27}kg/u}&=&7.73 *10^5m^2/kg\\
  \hbox{Silicium Si: }\frac{\sigma_{Si}}{28u*1.66*10^{-27}kg/u}&=& 8.18*10^5m^2/kg
  \end{eqnarray}


  
In a dust grain say the atoms will typically shadow each other and thus this
ratio ``the inverse darkness'' will be smaller than if the atoms were all
exposed to the collision considered. If we denote the average
number of atoms lying in the shadow of one atom by ``numberthickness'' we
will have for the ratio for the full grain say
\begin{eqnarray}
  \frac{\sigma}{M}|_{grain}&=& \frac{\frac{\sigma}{M}|_{atom}}
       {\hbox{``numberthickness''}}. \label{shadow}
\end{eqnarray}

  
If we insert in the grain a mass-wise dominating bubble, the whole object
will of course get a small ratio due to the higher mass,
\begin{eqnarray}
  \frac{\sigma}{M}|_{composed} &=& \frac{\sigma}{M}|_{grain}*\frac{M_{grain}}{M},
\end{eqnarray}
where $M$ is the mass of the bubble or if it dominates the whole composed
object, the dark matter particle.
  
On the average of course the ratio $\frac{M_{grain}}{M}$ of the dust
around the bubble and the bubble itself can never be bigger than the ratio of the
amount of dust-suitable mass to dark matter in the universe. So
noting that the grain should largely be made by the elements
heavier than helium, the so called ``metals'', and that these make up
only of the order of 1 \% of the ordinary matter which again is only
about 1/6 of the mass of the dark matter, we must have
\begin{eqnarray}
  \frac{M_{grain}}{M}&\le & 1\% /6 = 1/600. \label{accessible}
  \end{eqnarray}
But really of course not all the ``metal'' has even reached out to the
intergalactic medium, let alone been caught up by the
dark matter. So we expect an appreciably smaller value for this
ratio of dust caught by dark matter relative to the dark matter itself.

  
In earlier papers we have already used the dark matter
self-interaction in the low velocity limit extracted from Correa's fit to the dwarf galaxy data shown in Figure \ref{Correa} to give:
\begin{eqnarray}
  \frac{\sigma}{M}|_{v--> 0} &=& 15 m^2/kg.
\end{eqnarray}

We now wish to crudely estimate the amount of dust that might pile up around a dark matter bubble with a given velocity during the evolution of the Universe. There are two important effects to be taken into account. First of all the metal density was higher in the past due to the reduction in the ``radius" of the Universe by a factor $(1+z)^{-1}$ where $z$ is the red shift. Secondly the metallicity was lower in the past and we use the linear fits of De Cia et al. \cite{met} to its z dependence in our estimate of the rate of collection of metals by our pearls. 
We found that the most important time for the rate of collection of metals corresponds to $z = 3.3$, when the age of the Universe was 1.52 milliard years. At this time the rate of collecting metals for a given velocity was about 8.4 times bigger than if using the present metallicity and density.

So we might crudely estimate the amount of dust being collected
by an 8.4 times bigger density of metals than today in the 8.9 times younger
Universe, giving {\em effective numbers} for the dust settling:
\begin{eqnarray}
  \hbox{``metal density''}_{eff} &=&  3.71*10^{-31}kg/m^3*8.4\\
  &=& 3.1*10^{-30}kg/m^3\\
  &=&1.7*10^{-3}GeV/c^2/m^3.
  \end{eqnarray}

For orientation we could first ask how much metal-matter at all could be
collected by a dust grain while already of the order of $10^{-7}m$ in size, meaning
a cross section of $10^{-14}m^2$ and with a velocity of say 300 km/s =
$3*10^5 m/s$ during an effective age of the Universe of
1.52 milliard years =  
$4.8*10^{16}s$. We obtain

\begin{eqnarray}
  \hbox{``available metals''}&=&
  3*10^5m/s *4.8*10^{16}s*10^{-14}m^2*3.1*10^{-30}kg/m^3 \nonumber\\
  &=&4.4*10^{-22}kg \\
  &=&2.4*10^5 GeV,
  \end{eqnarray}
which is to be compared to what the mass of a $(10^{-7}m)^3$ large dust particle
with say specific weight $1000kg/m^3$ would be, namely
$10^{-18}kg$.

So such a ``normal'' size dust grain could not collect itself
in the average conditions in the Universe.

However if the grain to be constructed had lower dimension than 3, then the cross section could be larger for the same hoped
for volume and thus mass. Decreasing say the thickness in one of the dimensions
from the $10^{-7}m$ to atomic size $10^{-10}m$ would for the same
collection of matter give a 1000 times smaller mass. This
would bring such a ``normal size'' grain close to being just collectable in the average conditions in the Universe.

Our speculated stronger forces than usual due to the big homolumo gap
would not help much, because the grain cannot catch the atoms in 
intergalactic space which it does not come near enough to touch.

  
We shall now estimate the inverse darkness for such a dust grain of
dimension 2 or less attached to a dark matter bubble. In this case there is
no shadowing of the dust atoms and the parameter ``numberthickness" in
equation (\ref{shadow}) becomes unity. Also we 
  estimated that in the main period when the dust attached itself to the
  dark matter bubbles, we had $z=3.3$ and the age of the Universe was 1.52
  milliard years. The density of ``metals'' at that time was a factor
  $10^{-1}$
  times the one today. So the factor $1/600$ in equation (\ref{accessible}) for
  the  ``metals'' accessible to be caught by the
  dark matter composite particle becomes
  \begin{eqnarray}
    \frac{M_{grain}}{M} &=& 1\%/6 /10 = \frac{1}{6000}.
\end{eqnarray}
  So taking $\frac{\sigma}{M}|_{atom} = 7*10^5m^2/kg$ for the atoms of dust,
  we obtain our estimate for the inverse darkness of the dark matter particle
  composed with a dust grain of dimension 2 or less
  \begin{eqnarray}
    \frac{\sigma}{M}|_{composed} &=&
    \frac{\sigma}{M}|_{grain}*\frac{M_{grain}}{M}\\
    &=& 7*10^5m^2/kg /6000 \\
    &=& 1.2*10^2m^2/kg.
  \end{eqnarray}

  Our expected ratio
  \begin{eqnarray}
     \frac{\sigma}{M}|_{composed} &=& 120 m^2/kg
    \end{eqnarray}
should be compared with the value extracted from the dwarf galaxy data
     \begin{eqnarray}
    \frac{\sigma}{M}|_{Correa, v\rightarrow 0}&=&
    15 m^2/kg. 
    \end{eqnarray}
  
\subsection{Size of Individual Dark Matter Particles}


  In the approximation of only gravitational interaction of dark matter it is
  well-known that only the {\em mass density} matters, whereas the number
  density or the {\em mass per particle is not observable.}

  With other than gravitational interactions one could hope that it
  would be possible to extract from the fits in say our model, what the
  particle size should be. But the possibility
  for that in our model is remarkably bad! The Correa measurement yields just the ``inverse
  darkness'' ratio
  \begin{eqnarray}
    \frac{\sigma}{M}&=& \frac{\hbox{``cross section''}}{\hbox{mass}} 
    \end{eqnarray}

  Our estimate for the rate of 3.5 keV radiation from dark matter seen by DAMA - very crudely - was based on:
  \begin{itemize}
  \item The total kinetic energy of the dark matter hitting the Earth
    per $m^2$ per $s$ (but not on how many particles).
  \item The main part of that energy goes into 3.5 keV radiation of
    electrons.
  \item Estimate of a ``suppression'' factor for how small a part of this
    electron radiation comes from sufficiently long living excitations to
    survive down to 1400 m into the Earth.
  \end{itemize}

  None of this depends in our estimate on the size of the dark matter particles (provided it lies inside
  a very broad range)!


  If the dark matter particles were so heavy that the number density is so low
  that the observation over an area of about $1 m^2$ would not get an event through
  every year, then it would contradict the DAMA data.
  
  The rate of dark matter mass hitting a square meter of the Earth is
  \begin{eqnarray}
    \hbox{Rate} &=& 300km/s *0.3 GeV/cm^3\\
    &=& 3*10^5m/s *5.34*10^{-22}kg/m^3\\
    &=& 1.6*10^{-16}kg/m^2/s\\
    &=&5*10^{-9}kg/m^2/y
  \end{eqnarray}
  Taking the DAMA area of observation $\sim 1m^2$ we need to get more than
  one passage per year and thus
  \begin{eqnarray}
    M &\le & 5*10^{-9}kg\\
    &=& 3*10^{18}GeV.
    \end{eqnarray}
 Using the bubble internal mass density as estimated from the 3.5 keV homolumo gap, this upper bound implies that  
the bubble radius $R\le 10^{-7}m$.


 \begin{figure} 
  \includegraphics[scale=0.5]{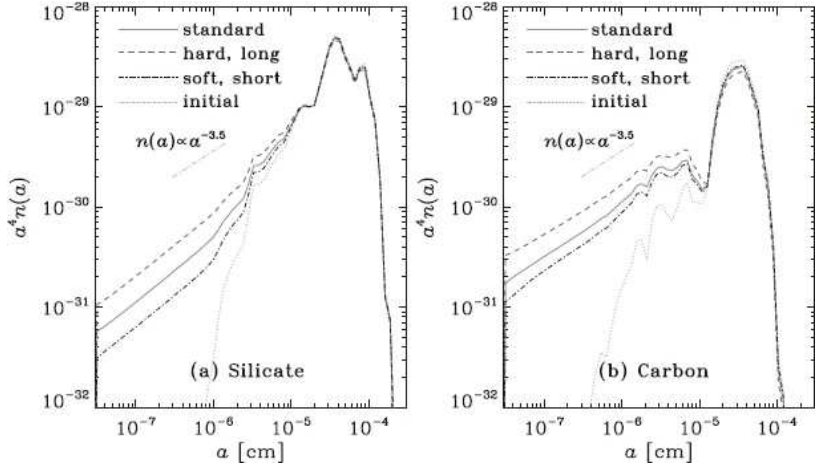}
 \caption{\label{size} Simulated size distribution for dust grains.}  
\end{figure}
  

  We can assume a typical grain size (see Figure \ref{size}) of $10^{-7}m$, say.
  Then using the low velocity limit $\frac{\sigma}{M} =15 m^2/kg$
  gives
  \begin{eqnarray}
    M&=& (10^{-7}m)^2/(15 m^2/kg)\\
    &=& 7*10^{-16}kg.
    \end{eqnarray}
  But if now the dust grain is less than 2 dimensional, the
  area for a grain with the same weight as a massive 3 dimensional one would be more than $\frac{10^{-7}m}{10^{-10}m} =1000$ times bigger, i.e an area bigger than 
  $10^{-11}m^2$. Correcting for this would give a bigger mass $M \ge 7*10^{-13}kg$.
  

\section{Conclusion}

We have reviewed and updated our dark matter model in which the dark matter
consists of bubbles of a speculated new phase of the vacuum, in which there has
collected so much ``ordinary'' matter that the surface tension of the
separation surface between the two types of vacuum can be spanned out. These
pearls are here assumed to be surrounded by a lower than three dimensional
grain of dust mainly made from atoms of higher atomic weight than hydrogen and helium.

We suppose that the Hausdorff dimension of the grain of dust is so low that
the interaction between the pearls with their dust corresponds to effectively having
no shadowing of the grain atoms by each other (with added up dimensionality
less than 2). We used the general chemical abundances and estimated a low velocity 
inverse darkness of $120 m^2/kg$ for our pearls. This is only one order of magnitude
larger - thus it essentially agrees with - the value $15 m^2/kg$ found by Correa \cite{CAC}. This is summarized in Table \ref{suc}  as point 1.

In item 2 in the table we see that the estimate for the value $v_0$ of the velocity at which
the inverse darkness falls significantly down as a function of the velocity
is $0.7 cm/s$ if we do not take the hardening of the dust grain seriously,
while it is $77 km/s$ if we do take this hardening seriously. From the Correa estimate
using the dwarf galaxy data one finds $v_0=220 km/s$, so only the estimate taking the hardening seriously agrees with experiment.

The rest of the items in Table \ref{suc} are other order of magnitude
estimates checking the viability of our model. Thus    
item 3 estimates the rate of events in the DAMA-LIBRA experiment formulated
in terms of the quantity $suppression$, which denotes the fraction of the excitations made in the dark matter pearls on entering the Earth, which survive down until the pearl reaches the detector.

The similar item 4 for the XENON1T experiment is now obsolete in as far as
the effect found in this experiment was not reproduced after the
radon gas was better cleaned away in XENONnT, so it was probably  $\beta$ decay of $^{214}Pb$ that was responsible for the previous effect.

Item 5 called ``Jeltema'' represents the very
strange observation of the $3.5keV$-line from the Tycho Brahe supernova remnant,
which should not have enough dark matter to produce the $3.5 keV$-line so as
to be observed at all. But due to our dark matter particles being excitable
by the large amount of cosmic rays in the supernova remnant, we indeed
could get agreement with the observed rate of $2.2*10^{-5}photons /cm^2$ coming
from the supernova remnant.
 
As item 6 we list the fit of the overall factor in the fit by
Cline and Frey to the $3.5 keV$-line sources together with the
very energy $3.5keV$ by one combination of our parameters for the
model $\frac{\xi_{fS}^{1/4}}{\Delta V}$. Actually this combination is essentially the Fermi momentum of the electrons in the highly compressed matter in the interior of our our bubbles. This fitting is only sensitive to the density of the matter inside the pearls and does not depend on the size of the pearls at
the end. So this successful fit actually originates from earlier articles on our
model, when we considered the pearls to be cm-sized and so
heavy that an impact in Tunguska could have caused a major catastrophe \cite{Tunguska}.

The last item, item 7, just reviews the fact that we found approximately the same
 $3.5 keV$ at first in three different places. However now after the sad development for our model in the recent XENONnT experiment \cite{XENONnT} only in two places, namely in the
satellite etc. observations of the 3.5 keV X-ray line and in the average energy of the modulating part of the DAMA-LIBRA observed events.

In Table \ref{bounds} we present some information on the mass of the single
dark matter particle mass $M$ (supposedly dominating the mass of the
dust grain).


  
  \begin{table}
  \caption{\label{suc} Successes}
    \begin{tabular}{|c|c|c|c|c|
      }
  \hline
  \# \& exp/th & Quantity &value  &related Q. & value
  \\
  \hline
  1. & Dwarf Galaxies &&&\\
  exp & inv. darkness =& $15m^2/kg$& $\frac{M_{grain}}{M}$& $2*10^{-5}$\\
  th&= $\frac{\sigma}{M}|_{v\rightarrow 0}$ &$120 m^2/kg$ &&$1.6*10^{-4}$\\
  \hline
  2.& Dwarf Galaxies&&&
  \\
  exp &Velocity par. $v_0$ & $220km/s$
  & $4r_{dust}E$ &
  $8.1*10^{13}kg/s^2$
  \\
  th. & with hardening &$77 km/s$&$4r_{dust}E$    &$1*10^{13}kg/s^2$
  \\
  th. & without hard.& $0.7 cm/s$&$4r_{dust}E$&$400 kg/s^2$
  \\
  \hline
  3.& DAMA-LIBRA&&&
  \\
  exp &
  &$0.041 cpd/kg$ &suppression
  &
  $1.6*10^{-10}$
  \\
  th& air
  &$
  0.16 cpd/kg$&&$
  6*10^{-10}$
  \\
  th&stone & $1.6*10^{-5}cpd/kg$ &&$6*10^{-14}$
  \\
  \hline
  4.&Xenon1T&&&
  \\
  exp&
  & $
  2*10^{-4} cpd/kg$&suppression
  &$
  6*10^{-13}$
  \\
  th & air &$
  0.16 cpd/kg$&&$
  6*10^{-10}$\\
  th& stone &$1.6*10^{-5}cpd/kg$&&$6*10^{-14}$
  \\
  \hline
  


  5.&Jeltema \&
  P.&&&
  \\
  exp&counting rate&$2.2 *10^{-5}phs/cm^2/s$&$\frac{\sigma}{M}|_{Tycho}$
  &$5.6*10^{-3}cm^2/kg$
  \\
  th&&$3*10^{-6}phs/s/cm^2$&$1\% *\alpha*\frac{\sigma}{M}|_{nuclear}$
  &$8*10^{-4}cm^2/kg$
  \\
  \hline
  6.&Intensity 3.5 kev&&&
  \\
  exp& $\frac{N\sigma}{M^2}$ & $10^{23}cm^2/kg^2$
  &$\frac{\xi_{fS}^{1/4}}{\Delta V}$&$0.6 MeV^{-1}$
  \\
  th& & $3.6*10^{22}cm^2/kg^2$&&$0.5 MeV^{-1}$
  \\
  \hline
  7.& Three Energies&&&
  \\
  ast& line & 3. 5 keV&&
  \\
  DAMA& av. en.& 3.4 keV&&
  \\
  Xen.&av. en. &3.7 keV&&
  \\
  \hline
 \end{tabular}
  \end{table}

 \newpage 
  
  
\begin{center}
  \begin{table}
    \caption{\label{bounds} Mass $M$ bounds and estimates} 
    \begin{tabular}{|c|c|c|c|c|
    }
  \hline
  Description&$R$&$\Delta R$&$M$&$\Delta M$
  \\
  \hline
  Faster than year&$\ge 1.0*10^{-9}m$&&$\ge 2.1*10^{-15}kg$&
  \\
  Corrected
  & $\ge 3.1*10^{-9}m$&&$\ge 6.5*10^{-14}kg$&
  \\
  \hline
  Dust enough&$\ge 1.0*10^{-9}m$&&$\ge 2*10^{-15}kg$&
  \\
  \hline
  Velocity dep.& $\approx 10^{-8}m$&big&$\approx 10^{-13}kg$
  &big
  \\
  w. E= $400^4$&$10^{-10}m$&&$\approx 2*10^{-18}kg$
  &
  \\
  \hline
  DAMA stream&$\le 10^{-7}m$&& $\le 5*10^{-9}kg$&\\
  \hline
  Grain size $10^{-7}m$&$7*10^{-10}m$ & & $7*10^{-16}kg$&\\
  \hline
  \end{tabular}
  
  \end{table}
  \end{center}


\section*{Acknowledgements}
We acknowledge our status as emeriti at respectively Glasgow University
and the Niels Bohr Institute, and H. B. N. acknowledges discussions at conferences like of cause the Bled Workshop but also at Corfu.


\end{document}